\documentclass{aastex}
\usepackage{graphicx}
\input psfig.sty
\usepackage{multirow}
\usepackage{textcomp}
\usepackage[english]{babel}
\usepackage{natbib}
\usepackage{amsmath,amssymb,mathrsfs}
\graphicspath{{figure/}}
\usepackage[top=2.5cm, inner=2.5cm, outer=3cm, bindingoffset=0cm]{geometry}
\def \1856{RX J1856}

\begin{document}
%%%%%%%%%%%%%%%%%% END OF PREAMBULE %%%%%%%%%%%%%%%%%%
\shorttitle{A Comment on ``}
\shortauthors{Turolla et al.}

\title{A Comment on ``A note on polarized light from Magnetars: QED effects and axion-like particles'' by
L.M. Capparelli, L. Maiani and A.D. Polosa}

\author{R. Turolla\altaffilmark{1,2}, S. Zane\altaffilmark{2}, R. Taverna\altaffilmark{1}, D. Gonz\'alez Caniulef\altaffilmark{2}, R.P. Mignani\altaffilmark{3,4}, V. Testa\altaffilmark{5}, K. Wu\altaffilmark{2}}
\altaffiltext{1}{Department of Physics and Astronomy,
University of Padova, Via Marzolo 8, I-35131 Padova, Italy}
\altaffiltext{2}{Mullard Space Science Laboratory, University College London, Holmbury St. Mary, Dorking, Surrey, RH5 6NT, UK}
\altaffiltext{3}{INAF-Istituto di Astrofisica Spaziale e Fisica Cosmica Milano, via E. Bassini 15, 20133, Milano, Italy}
\altaffiltext{4}{Janusz Gil Institute of Astronomy, University of Zielona G\`ora, Lubuska 2, 65-265, Zielona G\`ora, Poland}
\altaffiltext{5}{INAF-Astronomical Observatory of Rome, via Frascati 33, I-00078, Monte Porzio Catone, Italy}

\begin{abstract}

The recent detection of a large polarization degree in the optical emission of an isolated neutron star led 
to the suggestion that this has been the first evidence of vacuum polarization in a strong magnetic field, an effect predicted by quantum 
electrodynamics but never observed before. This claim was challanged in a paper by  \cite{capparelli17}, according to whom
a much higher polarization degree would be necessary to positively identify vacuum polarization. Here we show that 
their conclusions are biased by several inadequate assumptions  and have no impact on the original claim.

\end{abstract}


\section{Introduction}\label{intro}

In a recent paper, \cite{capparelli17} presented some considerations in connection with
the discovery of a relatively high linear polarization degree ($\sim 16\%$) in the
optical emission of the isolated neutron star (NS) RX J1856.5-3754 \cite[hereafter \1856;][]{mignani17}.  Specifically, they expressed some 
criticism on the interpretation of the measured polarization  as the first evidence for vacuum birefringence, as stated by \cite{mignani17}.

According to
quantum electrodynamics (QED), photons propagate in a strongly magnetized vacuum in two normal polarization modes,
the ordinary (O) and extraordinary (X) one, with different refractive indices and this strongly influences the polarization
properties of the observed radiation \cite[e.g.][]{heyl00,heyl02,hardinglai06,taverna15,denis16}. This effect has been
searched for but never observed in terrestrial laboratories \cite[e.g the PVLAS experiment;][]{dellavalle16}.
The main point raised by \cite{capparelli17} is that the observed polarization degree in \1856\ would be not high enough
to provide an unambiguous signature of vacuum birefringence. According to their calculation, in fact, the maximum value of the
polarization degree, neglecting vacuum birefringence, is $\sim 40\%$, while accounting for QED effects it should be close to $100\%$.
Hence their conclusion that the measured value is too small to support the presence of vacuum birefringence in the magnetosphere of \1856.

Here we show that their conclusions are incorrect because of the oversimplified treatment of the magnetic
field around the neutron star they
use. Besides, their approach in computing the phase-averaged polarization observables appears flawed
and the constraints set by the observed values of the star parameters,
the surface magnetic field and the pulsed fraction, are not accounted for.

\section{Magnetic field topology}

\begin{figure*}[htbp]
\begin{center}
\includegraphics[width=15cm]{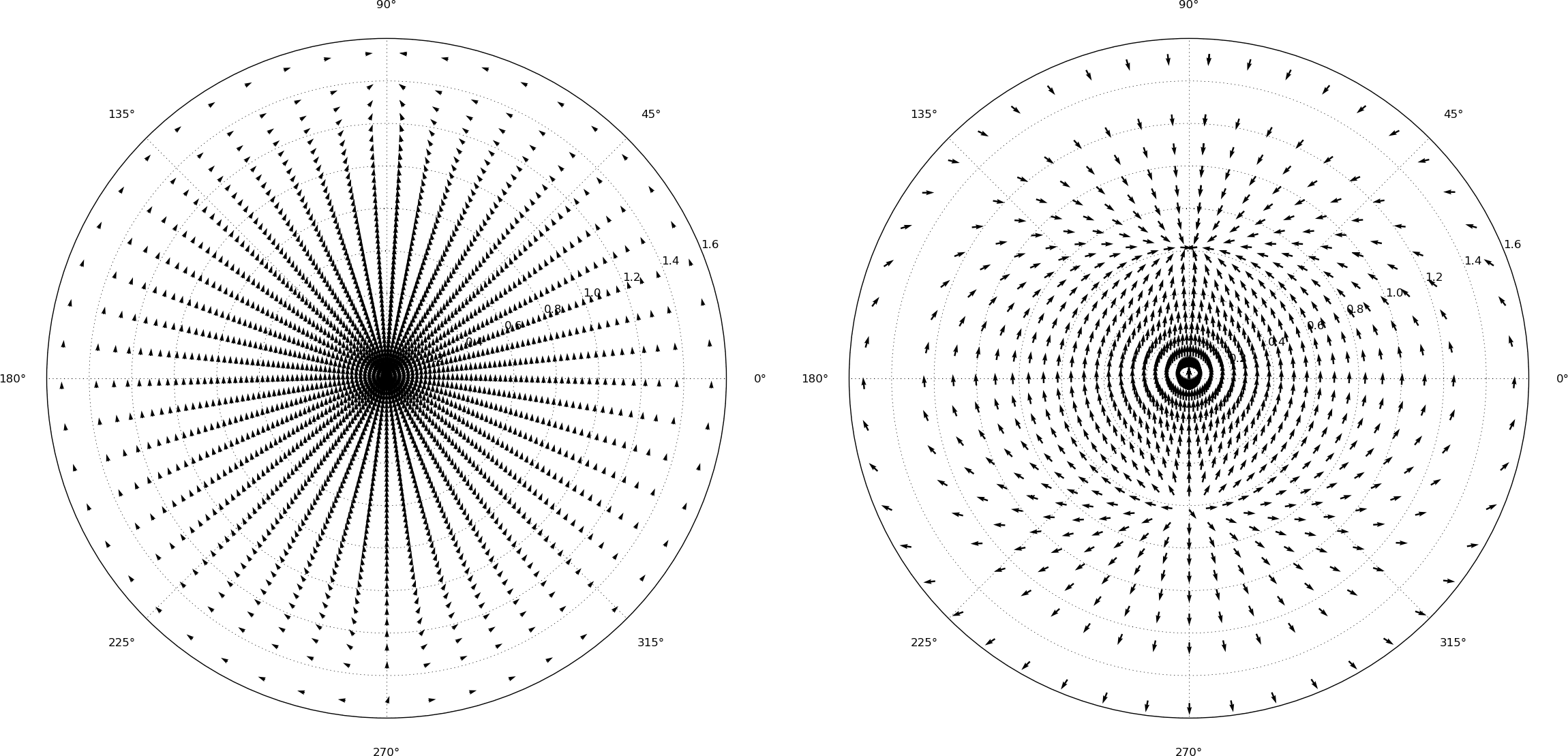}
\caption{Magnetic field distribution over the star surface projected on
the plane perpendicular to the LOS for both a meridional field (left-hand
panel) and a dipolar field (right-hand panel).}
\label{figure:arrows}
\end{center}
\end{figure*}

\cite{capparelli17} considered a simplified magnetic field configuration in which
the magnetic field vectors are tangent to the meridians of the star surface. This assumption
works well to illustrate how the photon electric field direction
changes as the radiation propagates in the magnetized vacuum around the
source (as shown in their figure 1). However, as also they themselves note, the
external field of an NS is most likely a core-centered dipole, and this seems indeed the case
for \1856, as recently discussed by \cite{popov17}. Actually, while meridional field lines
well approximate the dipolar ones far from the star, the agreement becomes
worse and  worse closer to the star surface, where radiation is emitted. In assessing the maximum observed value of the
polarization degree in the absence of vacuum birefringence (QED-off case), \cite{capparelli17} keep to this unrealistic approximation (see their footnote 2) and get
a value $\lesssim 40\%$ for photons initially $100\%$ polarized in the X mode. We stress that the magnetic field topology
is key in computing  the observable polarization signal (in the QED-off limit). Since the magnetic configuration assumed by \cite{capparelli17}
is much more uniform than the dipolar one, this produces an overestimate of the
polarization degree. This can be clearly seen in figure \ref{figure:arrows} which shows
the magnetic field over the star surface (projected
on the plane perpendicular to the line of sight, LOS) for both magnetic configurations.
The meridional magnetic field configuration is indeed more organized/uniform than that of the dipolar field,
so that the expected polarization degree in the absence of QED effects is
higher for the meridional field. 

This can be explained by the fact that the Stokes parameters associated to each
photon have to be rotated around the LOS in order to be all referred to the same frame (the
polarimeter frame) before they are summed together to obtain the overall polarization signal \cite[see][for further details]{taverna15}. 
Since the rotation angle $\alpha$ is indeed the angle between the projection of the local $B$-field perpendicular to the LOS at the emission point
and the polarimeter reference axis, it becomes clear that the configuration of the magnetic field on the star
surface strongly influences the observed polarization degree when QED effects are neglected. In particular a 
narrow range of variation for the angle $\alpha$ over the star surface translates into a larger polarization degree
at the observer. 

\begin{figure*}[htbp]
\begin{center}
\includegraphics[width=15cm]{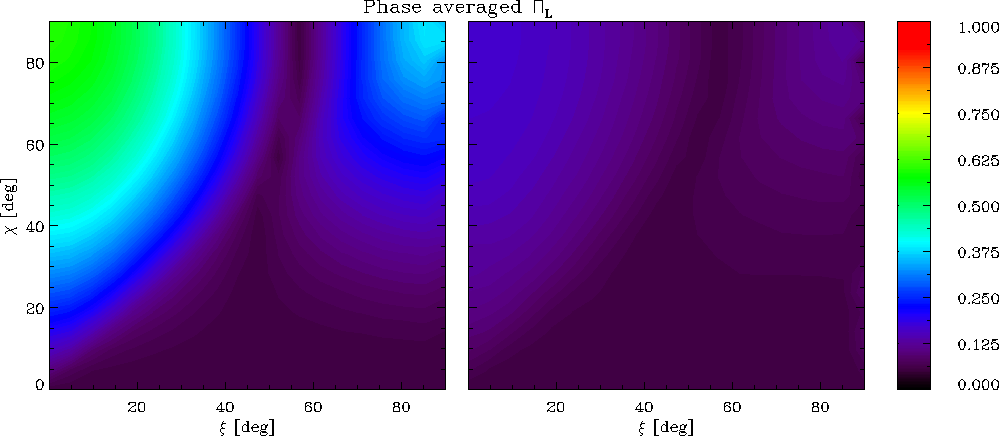}
\caption{Contour plot of the phase-averaged polarization degree $\Pi_\mathrm{L}$ (see text)
in the absence of QED effects for the meridional (left-hand panel)
and dipolar (right-hand panel) magnetic field configurations,  as
a function of the geometrical angles $\chi$ (between the LOS and the rotation
axis) and $\xi$ (between the magnetic and the rotation axes).}
\label{figure:contour}
\end{center}
\end{figure*}

To better assess this point, we recomputed the polarization observables with our ray-tracing code using a setup
similar to that adopted by  \cite{capparelli17}, i.e. meridional field and $100\%$ polarized blackbody photons, obtaining a value of the polarization degree
in the QED-off limit of $\sim 50\%$ for the most favourable viewing geometry, much higher than that for a dipolar field, $\sim 13\%$ 
for the same photon input. This is clearly illustrated in figure
\ref{figure:contour} which shows the observed phase-averaged polarization degree $\Pi_\mathrm L$ as a function of the two geometrical angles $\xi$ and $\chi$, the magnetic axis and the LOS, respectively, make with the rotation axis.

The value of $\sim 40\%$ reported by \cite{capparelli17} as the upper limit for the observed polarization degree in the absence of QED effects is strongly affected by their
oversimplified treatment of the star magnetic field, has no physical basis, and cannot be compared with the
results discussed in \cite{mignani17}. As a consequence, their statement that the measure of $\Pi_\mathrm L\sim 16\%$ in \1856 is insufficient to infer vacuum birefringence is devoid of any meaning.

\section{Phase-average of the Stokes parameters}

The definition adopted by \cite{capparelli17} for the polarization
degree is unclear. The polarization degree considered in
\cite{mignani17} is defined in terms of the (normalized) Stokes
parameters $Q$ and $U$ as $\Pi_\mathrm{L}=\sqrt{Q^2+U^2}$. The
FORS2 instrument used in the polarimetric observation measures the
Stokes parameters of the collected radiation. Due to the limit on the time resolution imposed by the exposure duration, 
only an average of the polarized signal over the rotational phase can be obtained. Since
the Stokes parameters are additive, the phase-averaged
polarization degree can be obtained only by averaging the Stokes
parameters before computing the polarization observables, a fact \cite{capparelli17} completely neglect. Indeed,
the effect of phase-averaging over the star rotation is 
that of an effective depolarization, and this holds also when QED
effects are accounted for \cite[as shown in
][]{taverna15,denis16}. As a key example, one may consider the
configuration shown in figure 1 of \cite{capparelli17}, and set
the star spin axis along the LOS. After a full rotation of the
star, a polarimeter collects photons with electric field pointing
in different directions. This reduces the polarization degree down
to zero (even using the simplified meridional configuration for
the magnetic field), despite the fact that a phase-resolved
polarization measurement gives, instead, $100\%$. This means
that the phase-averaged polarization degree as measured at infinity strongly 
depends on the viewing geometry and the observed signal in the presence of vacuum birefringence is not necessary $100\%$
polarized. In particular, it attains a maximum when 
the phase-averaging effects are less important, i.e. in the case of an orthogonal 
rotator seen perpendicularly to the rotation axis. 
%On the other hand, the effects of the 
%neutron star rotation are completely neglected in the calculation by \cite{capparelli17}. 

\section{Constraints on the star viewing geometry and magnetic field}

In order to observe the maximum polarization degree
\cite{capparelli17} consider the case of the LOS perpendicular
to the star magnetic axis.
%the most favourable viewing geometry,ry
%i.e. that with the LOS perpendicular to the star magnetic axis. 
We point out that, actually, this is not entirely true. In fact, the
angle between the LOS and the magnetic axis depend in general on
the rotational phase \cite[][]{taverna15,denis16}. Hence, as illustrated 
in the previous section, the only geometrical configuration which allows 
to observe the maximum polarization degree turns out to be that in which 
both the LOS and the magnetic axis are orthogonal to the spin axis of the star.
%As a consequence, due to the average over the rotational phase of the
%Stokes parameters, the only geometrical configuration in which the
%observed polarization degree attains a maximum turns out to be the
%case of an aligned rotator seen perpendicularly to the star
%spin/magnetic axis. 
This is also shown in the contour plots of figure \ref{figure:contour} above, 
as well as in those shown in \cite{mignani17}, where the maximum polarization 
degree is attained at the top-left corner.

However, observations place several constraints on the geometrical
angles $\chi$ and $\xi$ for \1856. In particular the pair
$(\chi\,, \xi)$ must be compatible with the observed pulsed
fraction in the X-rays \cite[$\sim 1.2\%$;][]{mereghetti07}. This
was accounted for in \cite{mignani17}. In addition, also on the basis of
spectral observations, a further constraint was placed by \citet{ho07},
forcing $\chi$ and $\xi$ to vary in narrow ranges, i.e. $\chi\approx 20^\circ$--
$45^\circ$ and $\xi\lesssim 6^\circ$. Including these constraints, it becomes
clear that the viewing configuration which gives the maximum polarization 
degree (i.e. $\chi=90^\circ$, $\xi=0^\circ$) is ruled out by observations.

\cite{capparelli17} note that a $16\%$ polarization degree
may indeed be the maximum value attainable in the case the surface radiation
is not $100\%$ polarized. Since in this situation the estimate of the polarization 
degree at the emission may be uncertain, they claim that ``only very high
degrees of linear polarization ($\gtrsim 50\%$) would be the indisputable 
footprint of QED birefringence effects''. However, \cite{mignani17} have
clearly shown that, once all the available observational constraints are accounted for, 
a $16\%$ polarization degree is indeed sufficient for a strong statement 
on the presence of QED effects even considering the worst case of 
surface blackbody radiation $100\%$ polarized in the extraordinary mode.

Furthermore, \cite{capparelli17} consider the source as a magnetar
candidate, with a surface magnetic field $B\sim 10^{14}$ G. This
is incorrect, since \1856\ belong to the neutron star class known
as the  XDINSs \cite[see e.g.][for a review]{turolla09}, and its spin-down magnetic field is $\sim 10^{13}\
\mathrm G$, one order of magnitude lower \cite[][]{vkk08}. Besides
the effects on the polarization observables, this may also impact
on their analysis on axion-like particle effects on the
polarization signal.

\section{Conclusions}

\cite{capparelli17} state ``Finally, even if we assume that every
single point of the star emits polarized light, then a degree of
polarization of $16\%$ may be reached in the absence of QED
effects just by a favourable orientation of  the magnetic axis of
the star with respect to the observation line, as is also evident
from some models  analyzed in \cite{mignani17}.'' However, by performing a
careful analysis (i.e. taking a dipolar magnetic field on the star
surface, computing the polarization observables from
phase-averaged Stokes parameters and accounting for the
geometrical constraints given by the observations), we showed that
the minimum polarization degree sufficient for a conclusive claim
about vacuum birefringence effects for \1856\ is indeed much lower
than the maximum polarization attainable, which is not $40\%$ but
rather $\lesssim 14\%$. Hence, the claim put forward by
\cite{capparelli17} is totally unjustified.

\acknowledgements

DGC acknowledges a Becas-Chile CONICYT Fellowship (No. 72150555).

%Acknowledge anyone ?
%
%

\end{document}